# Toward a Comprehensive Model of Snow Crystal Growth:
# 10. On the Molecular Dynamics of Structure Dependent Attachment Kinetics

Kenneth G. Libbrecht

Department of Physics
California Institute of Technology
Pasadena, California 91125
kgl@caltech.edu

**Abstract.** I examine the molecular dynamics of ice growth from water vapor, focusing on how the attachment kinetics can be augmented by edge-dependent surface diffusion. Although there are significant uncertainties in developing an accurate physical model of this process, it is possible to make some reasonable estimates of surface diffusion rates and admolecule density enhancements, derived from our basic understanding of ice-crystal growth processes. A quantitative model suggests that edge-dependent surface diffusion could substantially enhance terrace nucleation on narrow faceted surfaces, especially at the onset of surface premelting. This result supports our hypothesized mechanism for structure-dependent attachment kinetics, which readily explains the changes in snow crystal growth morphology with temperature depicted in the well-known Nakaya diagram. Many of the model features described here may be amenable to further quantitative investigation using existing computational models of the molecular structure and dynamics of the ice surface.

## ❄ Structure Dependent Attachment Kinetics

As the title suggests, this series of papers describes my ongoing efforts to develop a comprehensive physical model of the molecular dynamics of snow crystal growth. The need for a specific research effort in this direction first became apparent in the 1930s with the work of Nakaya and collaborators, who discovered a puzzling morphological progression of snow-crystal growth behavior with temperature [1954Nak]. Confirmed and extended by numerous researchers [1957Kob, 1958Hal, 1961Kob, 2004Bai, 2009Bai, 2021Lib], observations of snow crystal growth in air can be characterized (in broad terms) as being platelike above -3 C, columnar from -4 C to -6 C, platelike again from -11 C to -18 C, and columnar again below -30 C. These observations are often summarized in the *Nakaya diagram*, which depicts snow crystal growth morphology as a function of temperature and water vapor supersaturation. Developing a suitable physical model that explains these empirical observations has been a considerable scientific challenge for many decades [1958Hal, 1963Mas, 1982Kur, 1984Kur, 2001Nel, 2021Lib].

While the Nakaya diagram provides an accepted summary of snow crystal growth morphologies in normal air, observations of ice growth in low-pressure environments do not show similar transitions between platelike and columnar forms as a function of temperature [1969Fuk, 1972Lam, 1976Rya, 1982Bec2, 1983Bec, 1984Kur1, 1989Sei, 1998Nel, 2003Lib, 2021Lib]. For the specific case of growth near -14 C, I proposed that the attachment kinetics on faceted prism surfaces might be enhanced when the facet width becomes sufficiently narrow, as this could explain the discrepancy in growth rates between broad facets growing in a near-vacuum environment and thin plates growing in normal air [2003Lib1].

I referred to this phenomenon as structure-dependent attachment kinetics (SDAK), meaning that the *mesostructure* of the crystal morphology could alter the attachment kinetics that is typically determined entirely by the *nanostructure* of the crystal surface at molecular scales. As more and better growth measurements became available [2013Lib, 2021Lib], it became clearer that some form of SDAK phenomenon would likely be required to create a coherent explanation of the different data sets. However, finding a suitable physical mechanism with the necessary characteristics remained problematic.

I recently hypothesized that edge-dependent surface diffusion could provide the desired SDAK effect [2019Lib1], and I further developed this hypothesis into a comprehensive attachment kinetics (CAK) model for snow crystal growth [2020Lib1, 2020Lib2, 2021Lib]. The CAK model begins with the attachment coefficients on broad faceted surfaces, parameterized by

$$\alpha_x(\sigma_{surf}) = A_x e^{-\sigma_{0,x}/\sigma_{surf}} \qquad (1)$$

where $x$ stands for the principal *basal* or *prism* facets and $\sigma_{surf}$ is the water-vapor supersaturation at the ice surface. The nucleation parameter $\sigma_{0,x}$ derives from the terrace step energy at the ice/vapor interface, and $A_x$ depends on admolecule surface diffusion and other factors. This functional form is provided by terrace-nucleation theory [e.g., 1994Ven, 1996Sai, 1999Pim, 2002Mut], and the CAK model values of $\alpha_{basal}$, $\alpha_{prism}$, $A_{basal}$, and $A_{prism}$ as a function of temperature $T$ are shown in Figure 1 [2021Lib].

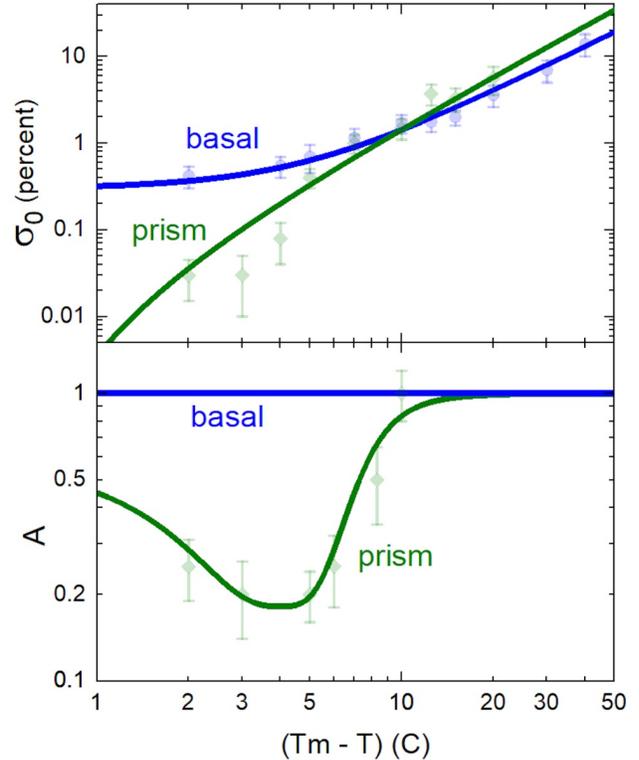

Figure 1: Model parameters for the growth of broad basal and prism facets as a function of temperature, as prescribed the CAK model. The parameters are defined in Equation 1, with $T_m \approx 0\,C$ being the melting point of ice. The nucleation parameter $\sigma_0$ derives from the terrace step energy at the ice/vapor interface, as dictated by classical nucleation theory. The $A$ parameter depends on admolecule surface transport and other factors that are difficult to determine accurately. These parameters were estimated from a variety of laboratory measurements of ice growth as a function of temperature and supersaturation, as described in [2021Lib].

The formation of platelike snow crystal morphologies above -3 C and columnar forms below -30 C follows immediately from the ratio

$\sigma_{0,basal}/\sigma_{0,prism}$ as a function of temperature. (The fact that $A_{prism} < 1$ at high temperatures provides only a small perturbation to this statement, and I will not address this feature of the CAK model in this paper, as the focus here is on prism growth near -14 C.) Although the functions $\sigma_{0,basal}(T)$ and $\sigma_{0,prism}(T)$ are only known empirically at present, they are related to the terrace step energies, which are fundamental equilibrium properties of the basal and prism surfaces. It appears likely that molecular dynamics simulations will soon yield accurate estimates of these functions [2020Llo], thereby providing a quantitative connection between snow crystal growth rates and *ab initio* calculations incorporating the quantum-mechanical properties of water molecular interactions.

One surprising prediction from the CAK model was that platelike morphologies should emerge near -5 C, in contrast to what had been reported in some versions of the Nakaya diagram [1961Kob, 1990Yok]. The model predicted that platelike morphologies at -5 C should appear both in vacuum and in air at low supersaturations, provided the crystals exhibit relatively broad basal and prism facets. This behavior was confirmed in targeted experiments [2012Kni, 2019Lib2], while columnar forms near -5 C generally emerge only when the basal facets are quite narrow.

More generally, the CAK model suggests that the SDAK effect on basal surfaces is necessary to produce columnar morphologies between -4 C to -6 C, while a similar phenomenon on prism surfaces is responsible for platelike forms between -11 C to -18 C. In both cases, the attachment kinetics is markedly different on narrow and broad facet surfaces.

The physical mechanism for the SDAK effect proposed in [2019Lib1] is outlined in Figure 2 for the case of platelike crystals growing near -14 C. As illustrated in Figure 2a, the model assumes that the equilibrium crystal shape of ice is nearly spherical, so surface energy considerations will round the edge of an established thin, platelike crystal growing slowly at a sufficiently low supersaturation. The large basal facets will not grow because of their substantial nucleation barriers, while the narrow prism edge will round to minimize the edge surface energy.

If the supersaturation is increased somewhat, the growth of both facets will be hindered by large nucleation barriers, and the corners will sharpen as illustrated in Figure 2b. In this quasi-static morphology exhibiting negligible growth, the corner radius of curvature will be roughly equal to

$$R \approx \frac{d_{sv}}{\sigma_{surf}} \qquad (2)$$

where $\sigma_{surf}$ is the supersaturation at the surface, $d_{sv} = \gamma_{sv}/c_{ice}kT \approx 1$ nm is the Gibbs-Thomson length, $\gamma_{sv} \approx 0.106$ J/m$^2$ is the solid/vapor surface energy, $c_{ice} = \rho_{ice}/m_{mol} \approx 3.1 \times 10^{28}$ m$^{-3}$ is the ice number density, and $kT \approx 3.6 \times 10^{-21}$ J at -14 C.

If surface diffusion is allowed on this nearly static crystal, as shown on the prism facet only in Figure 2c, then this process will tend to increase $R$, as this direction of molecular transport reduces the total surface energy. In this quasi-static picture, the nonzero supersaturation and anisotropic attachment kinetics create the nonequilibrium crystal shape in Figure 2b, and the resulting surface tension can drive the surface transport shown by the arrows in Figure 2c.

As described in [2019Lib1], the added surface transport increases the admolecule density on the narrow prism surface relative to what would normally exist on broad prism facets. This process does not change the terrace step energy, so does not directly change $\sigma_{0,prism}$. But the terrace nucleation rate does increase as a result, and this can be described mathematically as a lowering of the effective $\sigma_{0,prism}$ on the plate edge, yielding an SDAK effect.

According to the CAK model, surface diffusion is suppressed at low temperatures by a strong Ehrlich-Schwoebel barrier [1994Ven,

Figure 2: (right) How edge-dependent surface diffusion can produce the SDAK phenomenon on the thin prism edge of a platelike snow crystal [2019Lib1].

(a) When an established thin plate is held in near equilibrium conditions, the edge becomes rounded to minimize the edge surface energy.

(b) When the supersaturation is increased, the corners sharpen (Equation 2) from the Gibbs-Thomson effect, while high nucleation barriers prevent the facets from growing appreciably.

(c) Surface diffusion, if present, will tend to restore the equilibrium shape, depositing admolecules on the prism surface. The resulting increased admolecule surface density will enhance the normal nucleation rate, effectively lowering $\sigma_{0,prism}$. As described in the text, this process is especially significant at the onset of surface premelting on the prism surface, creating an "SDAK dip" near -14 C, as illustrated in Figure 3.

(d) As the QLL thickness increases and covers the prism facet, admolecule diffusion has a smaller effect on the nucleation dynamics, thus reducing the SDAK effect at temperatures above -14 C.

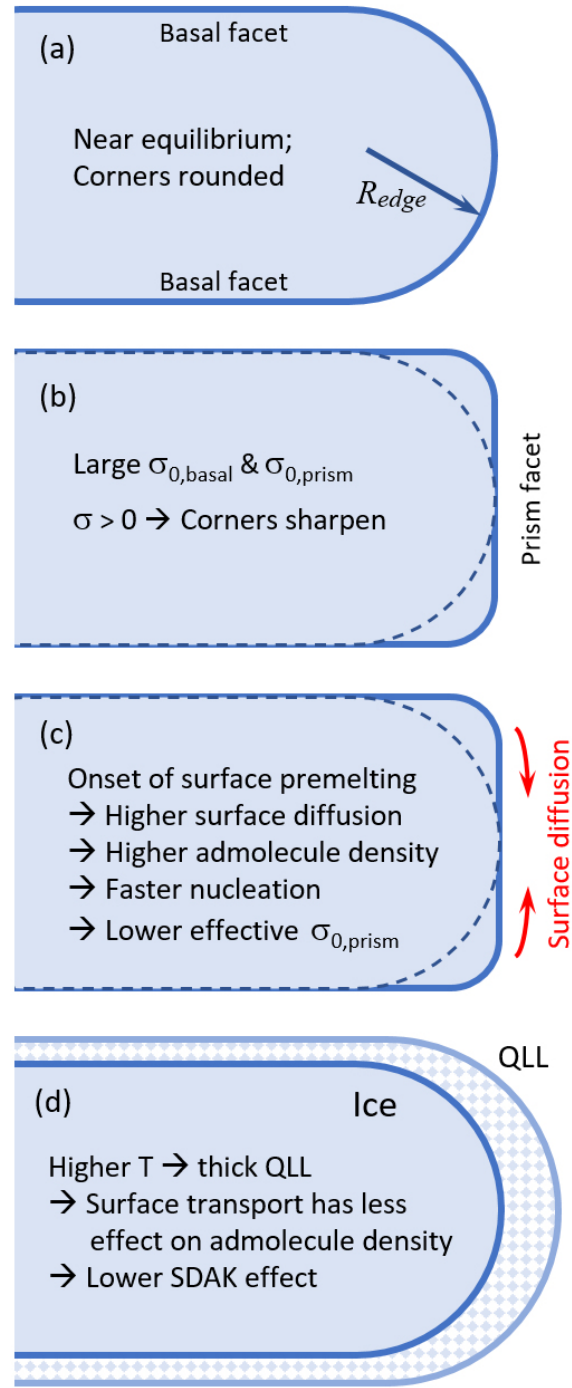

1996Sai, 1999Pim, 2002Mut], yielding a negligible SDAK effect at temperatures considerably below -14 C. This inhibition is greatly reduced at the onset of surface premelting, however, which the model postulates as occurring near -14 C on the prism facet. Near this temperature, therefore, surface diffusion is sufficient to substantially lower the effective $\sigma_{0,prism}$ on the narrow prism facet, yielding faster growth and thus facilitating the formation of thin, platelike structures.

At temperatures substantially higher than -14 C, surface premelting yields a sizeable quasi-liquid layer (QLL), as illustrated in Figure 2d, and I argue below that this results in a smaller perturbation in the effective $\sigma_{0,prism}$, and thus a reduced SDAK effect.

Putting all these pieces together yields an "SDAK dip" in the effective $\sigma_{0,prism}$ near -14 C, as illustrated in Figure 3. The CAK model additionally postulates a corresponding dip in $\sigma_{0,basal}$ near -4 C, and the combined effect yields columnar from -4 C to -6 C and platelike growth from -11 C to -18 C, as seen in the Nakaya diagram. Growth-rate measurements confirm the existence of SDAK dips at both -14 C [2020Lib1] and -4 C [2020Lib2], providing a compelling validation of the overall model structure. In its entirety, the CAK model presents a reasonable synthesis of known physical effects that can explain a broad range of experimental ice-growth data, including both morphological studies (summarized in the Nakaya diagram) and quantitative measurements of ice growth rates over a broad range of conditions.

My goal in the present paper is to examine the SDAK effect in more detail by developing a quantitative physical model of surface diffusion driven by surface tension near facet corners. While it is not possible to accurately describe all aspects of such a model, one can make some progress toward a better understanding of the SDAK phenomenon, identifying important physical parameters and suggesting additional targeted experimental and theoretical investigations. While unambiguous "proof" of the SDAK effect is a high bar for such a complicated phenomenon, this investigation finds that the model seems to be standing up well to closer scrutiny.

## ❋ Curvature Driven Surface Diffusion

In the SDAK model described above, surface diffusion driven by surface tension plays a critical role in facilitating the enhanced nucleation of new prism terraces on the thin edges of platelike crystals near -14 C, as illustrated in Figure 2. To explore and develop this SDAK model further, consider the sketch shown in Figure 4, showing one corner separating a basal and prism facet on a growing snow crystal.

I begin the discussion with the simplifying assumption that the ice/vapor surface energy $\gamma_{sv}$ is perfectly isotropic, meaning that the equilibrium crystal shape (ECS) of ice is spherical. While the ECS of ice in water vapor has not been definitively measured to date, the available evidence suggests that the surface energy is indeed nearly isotropic [2012Lib2]. If the ice/vapor ECS is faceted with sharp facet corners (possible, give the paucity of relevant experimental data, but unlikely), then the results presented in this paper would require major modification.

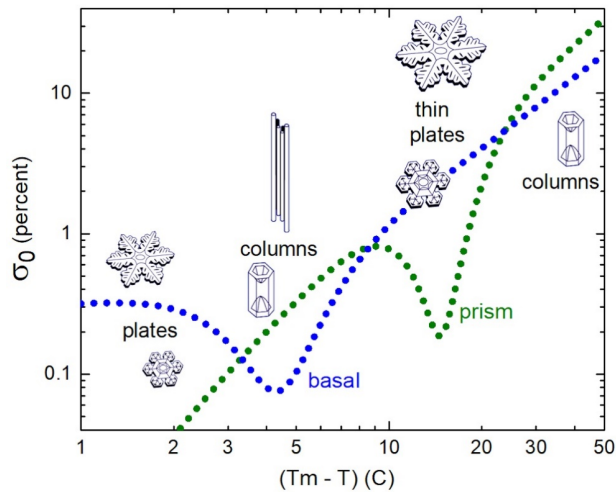

Figure 3: The CAK model incorporates the SDAK effect to add "SDAK dips" to the $\sigma_0$ plots in Figure 1, yielding the two curves shown here, which apply only to narrow facet structures. Together, these curves explain the platelike/columnar morphologies in Nakaya diagram as being strongly shaped by edge-dependent surface diffusion [2021Lib].

Given the assumption of isotropic surface energy, the change in admolecule binding energy with surface curvature is easily calculated from the Gibb-Thomson effect [1996Sai, 2021Lib]. For a sphere of radius $R_{sph}$ much larger than the molecular size, the binding energy at the curved surface is reduced (relative to a flat surface) by the amount

$$\delta E_{bind} = \frac{2\gamma_{sv}}{c_{ice}R_{sph}} = \frac{\gamma_{sv}}{c_{ice}}\kappa \qquad (3)$$

where $\kappa = 2/R_{sph}$ is the curvature of the spherical surface. For the corner geometry shown in Figure 4, the curvature is $\kappa = 1/R$, giving

$$\delta E_{bind} = \frac{\gamma_{sv}}{c_{ice}R} \qquad (4)$$

and Figure 5 shows a rough sketch of the admolecule binding energy as a function of surface position $x$ near the basal/prism facet corner.

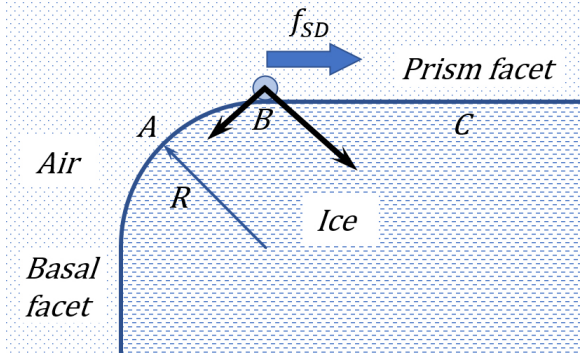

Figure 4: A schematic cross section of the corner between a basal and prism facet, where $R$ is the radius-of-curvature of the corner. Because of the asymmetry in the surface structure, an admolecule at position $B$ experiences a net lateral force $f_{SD}$ that drives surface diffusion toward the faceted prism surface. This model assumes an isotropic surface energy $\gamma_{sv}$, so the resulting admolecule surface transport tends to restore the crystal to its equilibrium spherical shape.

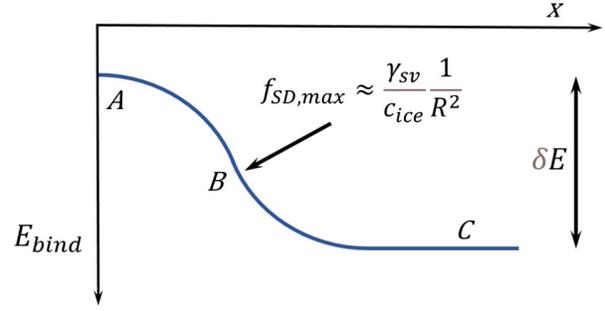

Figure 5: For the case of an isotropic surface energy $\gamma_{sv}$, the admolecule binding energy has the approximate form shown here as a function of surface position, where the points $A$, $B$, and $C$ correspond to the crystal locations shown in Figure 1. The lateral force $f_{SD}$ equals the spatial derivative of $E_{bind}$, and the maximum lateral force $f_{SD,max}$ occurs at point $B$, near the edge of the basal facet.

The lateral force $f_{SD}$ shown in Figure 4 is simply the derivative of $\delta E_{bind}$ with respect to surface position, and we see from Figure 5 that the maximum value of $f_{SD}$ is

$$f_{SD,max} \approx \frac{\gamma_{sv}}{c_{ice}} \frac{1}{R^2} \qquad (5)$$

If the surface supersaturation is $\sigma_{surf} \approx 1$ percent, Equation 2 gives a corner radius of $R \approx 100$ nm, so our next task is to examine how readily the surface-diffusion force $f_{SD}$ in Figure 4 drives admolecules from the corner region onto the prism facet. If the surface flux from this mechanism is high enough, then it may significantly increase terrace nucleation on the prism facet near -14 C, which is the central premise of the SDAK model presented above and in [2019Lib1].

## ❄ Admolecule Mobility

From statistical mechanics, we can write that diffusion in the presence of an external force produces an admolecule surface-diffusion drift velocity

$$\begin{aligned} v_{SD} &= \mu f_{SD} \\ &\approx \frac{D}{kT} f_{SD} \end{aligned} \qquad (6)$$

where $\mu$ is the mobility and $D$ is the admolecule diffusion constant. The second equality in this expression follows from the Einstein relation that connects mobility to the diffusion constant. For an admolecule placed at location $B$ in Figure 4, $f_{SD}$ can be replaced by the $f_{SD,max}$ given above.

The admolecule diffusion constant will likely depend strongly on the state of the underlying surface, being higher for smooth faceted surfaces than for rough surfaces. To my knowledge, $D$ on faceted surfaces is not well known from either experiments or theory, but we can make some rough approximations based on limiting cases.

The diffusion constant in liquid water is well known, and near 0 C we have $D_{water} \approx 1 \times 10^{-9}$ m²/sec [1999Pri]. Moreover, in the presence of well-developed surface premelting, surface diffusion takes place entirely in the quasi-liquid layer (QLL) at the ice surface, and molecular-dynamics (MD) simulations indicate that $D_{QLL}$ is roughly a factor of three less than $D_{water}$ [2011Gla]. At the opposite extreme, the diffusion constant in normal air is about $D_{air} \approx 2 \times 10^{-5}$ m$^2$/sec.

The diffusion constant is generally well approximated by $D \approx \ell v_{th}$, where $v_{th} \approx 350$ m/sec is a typical thermal molecular velocity near -14 C and $\ell$ is the mean-free-path between significant scattering events. The known liquid and air diffusion constants imply $\ell \approx 3$ pm in water and $\ell \approx 60$ nm in air.

On a faceted prism surface near -14 C, we expect a diffusion constant $D_{facet}$ somewhere between $D_{air}$ and $D_{QLL}$, but there is not much additional information we can use to narrow our estimate. On a faceted surface with a corrugated electronic potential, the phenomenon of Levy flights [2001Tso] can sometimes distort $D_{facet}$ to higher-than-expected values, because once an admolecule has enough energy to be above the corrugation scale, it may travel long distances before losing energy and dropping into a localized state. Given the substantial uncertainties, we choose $D_{facet} \approx 1 \times 10^{-7}$ m$^2$/sec for a faceted prism surface at -14 C, knowing that this value could easily be off by an order of magnitude or more. This estimate implies $\ell \approx 0.3$ nm on this surface, roughly equal to the ice lattice spacing $a \approx c_{ice}^{-1/3}$.

## ❄ Admolecule Dynamics

The admolecule surface density also factors into our model of the SDAK phenomenon, and again this quantity is not well known. Under equilibrium conditions, the flux of water molecules incident on an ice surface is $F_{in} \approx c_{sat} v_{th}$, where $c_{sat}$ is the water vapor number density above the surface and $v_{th}$ is an average of the thermal velocity distribution. The initial collision time is about $t_{coll} \approx a/v_{th} \approx 1$ psec, and we might expect many incident water molecules to experience a prompt reflection from the ice surface. Assuming some fraction $\lambda_s$ stick to the surface for a time substantially longer than $t_{coll}$, the effective incident flux becomes $F_{in,eff} \approx \lambda_s c_{sat} v_{th}$.

The flux of admolecules leaving the surface can be written $F_{out} = \rho/\tau$, where $\rho$ is the admolecule surface density and $\tau$ is the typical residence time on the surface [1996Sai]. Setting $F_{out} = F_{in,eff}$ yields a quasi-equilibrium admolecule surface density

$$\rho_0 \approx \lambda_s c_{sat} v_{th} \tau \qquad (7)$$

In this expression, only $c_{sat}$ and $v_{th}$ are well known, so $\rho$ and $\tau$ are quite uncertain. When an external supersaturation of $\sigma_{surf}$ is applied, the admolecule surface density increases to $\rho_0 + \delta\rho \approx \rho_0(1 + \sigma_{surf})$.

It is reasonable to assume a low occupation number of admolecules on a faceted surface, meaning $\rho_0 a^2 \ll 1$, and taking $c_{sat} \approx 5 \times 10^{22}$ $m^{-3}$ (the value at -14 C), $v_{th} \approx 350$ m/sec, and $\lambda_s \approx 0.1$ gives

$$\tau \ll \frac{1}{\lambda_s c_{sat} v_{th} a^2} \approx 10 \; \mu\text{sec} \qquad (8)$$

Our value for $\lambda_s$ is little more than a guess, but $\lambda_s = 1$ is unlikely (as it requires every incident molecule to initially stick to the faceted surface), and sputtering studies suggest that $\lambda_{s_s} < 0.01$ is also somewhat unlikely for a "soft" material like ice near the melting point. And, as with our earlier discussion, a rough guess is the best we can do at this point.

Another avenue for estimating $\tau$ comes from the equation $x_{SD} \approx \sqrt{D\tau}$ relating the surface diffusion length $x_{SD}$ to the diffusion constant and the residence time [1996Sai]. Unfortunately, the diffusion length is again difficult to determine accurately, and many measurements of $x_{SD}$ have been strongly affected by systematic errors associated with diffusion effects in air [2015Lib]. It appears than a rough estimate of $x_{SD} \approx 100$ nm is reasonable for a prism facet at -14 C, albeit with a large uncertainty. Note that $x_{SD}$ is likely to be substantially smaller on rough surfaces and will likely depend strongly on surface premelting as well. Combining $x_{SD} \approx 100$ nm with $D_{facet} \approx 1 \times 10^{-7}$ m²/sec gives $\tau \approx 10$ nsec, giving an admolecule occupation number of $\rho_{surf} a^2 \approx 10^{-3}$, and both these values seem reasonable as a first approximation.

Another way to look at the residence time is as a barrier penetration problem [1996Sai], which gives an estimate

$$\tau \approx \nu^{-1} \exp\left(\frac{E_{ad}}{kT}\right) \quad (9)$$

where $E_{ad}$ is the admolecule adsorption energy and $\nu$ is the lattice vibration frequency. Taking $\nu \approx 6$ THz from spectra showing translational lattice vibrations in ice [1969Bur] and $E_{ad} \approx E_{bind} \approx 8 \times 10^{-20}$ J/molecule gives $\tau \approx 1$ msec, which is clearly much too long. However, reducing $E_{bind}$ by a just factor of two gives $\tau \approx 10$ nsec, and this brief exercise is useful mainly as an illustration of how strongly $\tau$ depends on $E_{ad}$.

For the case of a one-molecule thick QLL, this means an effective admolecule occupation number of unity, and taking $\lambda_s \approx 1$ then gives $\tau \approx 1$ μsec. This long residence time is not unreasonable, as a QLL molecule is likely more tightly bound to the surface than an admolecule on a bare faceted surface. Using this value and estimating $D_{QLL} \approx 3 \times 10^{-10}$ m²/sec then gives $x_{SD} \approx 20$ nm in a QLL, again with a large uncertainty. Comparing these estimates with those above for a simple faceted surface, it appears that the diffusion constant is substantially lower, and the residence time is substantially higher, in the QLL. Moreover, the two factors cancel to some degree in the diffusion length, with the result that $x_{SD}$ does change by an extremely large factor with the addition of surface premelting.

## ❄ A Simple-surface Model

Another useful exercise is to consider a "simple-surface" model of admolecule diffusion near a corner, using the geometry in Figure 4 and assuming that a uniform surface with all relevant properties being independent of surface position $x$. Specifically, this basic model assumes essentially constant values of $x_{SD}$, $D$, and $\tau$ at all points on the surface, independent of surface structure. We can then write a one-dimensional surface-diffusion equation for the admolecule density $\rho$ as

$$\frac{\partial \rho}{\partial t} \approx D \frac{\partial^2 \rho}{\partial x^2} + \lambda_s c_{sat} v_{th} - \frac{\rho}{\tau} + \rho \mu \frac{\partial f_{SD}}{\partial x} \quad (10)$$

where the four terms on the right-hand side of this expression represent:

1) Normal diffusion governed by a simple surface diffusion constant $D$.
2) The effective deposition flux of admolecules from the vapor phase.
3) The admolecule sublimation flux from the surface.
4) Admolecule migration driven by surface curvature, described in conjunction with Figure 4.

Setting $f_{SD} = 0$, this equation has the steady-state solution $\rho = \rho_0$, as it must. Using $f_{SD}$ as described in Figure 5 and assuming that admolecule migration produces only a small perturbation with $\delta\rho/\rho_0 \ll 1$, we find that the main diffusion term is negligible, and the steady-state solution is approximately

$$\frac{\delta \rho}{\rho_0} \approx \frac{\partial f_{SD}}{\partial x} \frac{x_{SD}^2}{kT} \quad (11)$$

which has the approximate form illustrated in Figure 6. The maximum $\delta\rho/\rho$ occurs roughly a distance $R$ from point $B$ in Figure 4, with

$$\left(\frac{\delta\rho}{\rho_0}\right)_{max} \approx \frac{\gamma_{sv}}{c_{ice} kT} \frac{x_{SD}^2}{R^3}$$
$$\approx 0.01 \left(\frac{100 \text{ nm}}{R}\right)\left(\frac{x_{SD}}{R}\right)^2$$
$$\approx \sigma_{surf} \left(\frac{x_{SD}}{R}\right)^2 \quad (12)$$

where the last equality assumes a corner radius of $R \approx d_{sv}/\sigma_{surf}$.

If our estimate of $x_{SD} \approx 100$ nm is not too far off, then this perturbation to the normal admolecule surface density is substantial. When $\sigma_{surf} \approx 1$ percent and $R \approx 100$ nm, the nucleation dynamics will behave as if the supersaturation was double its normal value, meaning an effective $\sigma_{0,prism}$ of half its broad-facet value.

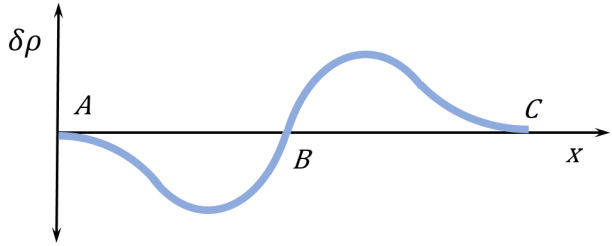

Figure 6: The "simple-surface" model described in the text yields a functional form of roughly this shape for the admolecule density perturbation, where the letters indicate the surface locations shown in Figure 4. The maximum value of the perturbation is estimated in Equation 12.

The principal result from this simple-surface model is that the effective change in admolecule density from curvature-driven surface diffusion could well be sufficient to substantially alter the nucleation dynamics on narrow prism facets near -14 C. A factor-of-two reduction in the effective $\sigma_{0,prism}$ is perhaps too small to reproduce the actual SDAK dip at -14 C, where $\sigma_{0,prism}$ is about one-tenth of its broad-facet value [2020Lib1].

Nevertheless, a factor of two decrease in $\sigma_{0,prism}$ is clearly not negligible. Setting aside its deficiencies and uncertainties, this rough calculation of $\delta\rho/\rho_0$ suggests that a curvature-driven surface-diffusion mechanism may well play an important role in the SDAK phenomenon, and in snow crystal growth dynamics more generally. We are encouraged, therefore, to consider next a more complex model to address the temperature dependence inherent in the SDAK dip.

## ❄ Terrace Step Dynamics

Although the simple-surface model in the previous section defines the essential physics of how surface-tension-driven surface diffusion can alter the attachment kinetics, this basic picture is too simplistic to explain the observed SDAK phenomenon. As illustrated in Figure 3 and quantified in laboratory measurements [2020Lib1, 2020Lib2], the SDAK dip on the prism facet is a remarkably sharp feature, greatly reducing the effective $\sigma_{0,prism}$ over a narrow temperature range centered at -14 C. Explaining this strong dependence on temperature requires that we examine the onset of surface premelting in detail.

The CAK model hypothesizes a transition to surface premelting with the general physical characteristics illustrated in Figure 7 [2019Lib1]. At low temperatures, with $T \ll -14$ C on the prism facet, we expect that surface premelting will be essentially absent, so the ice surface will be defined by a clean lattice structure with well-defined terrace steps. This is the standard picture for a crystalline solid (typically applied to materials with high melting points and low vapor pressures), and admolecule diffusion on such surfaces is often strongly hindered at terrace steps by the Ehrlich-Schwoebel barrier [1994Ven, 1996Sai, 1999Pim, 2002Mut], as indicated in Figure 7a. If this situation applies to ice, then surface diffusion will not transport admolecules to the top prism terrace in appreciable numbers, resulting in no SDAK effect when $T \ll -14$ C.

When the temperature approaches -14 C, however, then the CAK model hypothesizes that surface premelting will begin to alter the prism facet surface. As illustrated in Figure 7b, it stands to reason that the first effects of premelting will occur at terrace steps, as the molecular binding is weakest at these locations. The resulting amorphous structure "softens" the terrace step in that it greatly reduces the Ehrlich-Schwoebel barrier, thus permitting admolecule transport to the top prism facet. If the prism facet is sufficiently narrow, and the over-edge transport is driven by the surface-tension force illustrated in Figure 4, then the increased admolecule density will enhance the

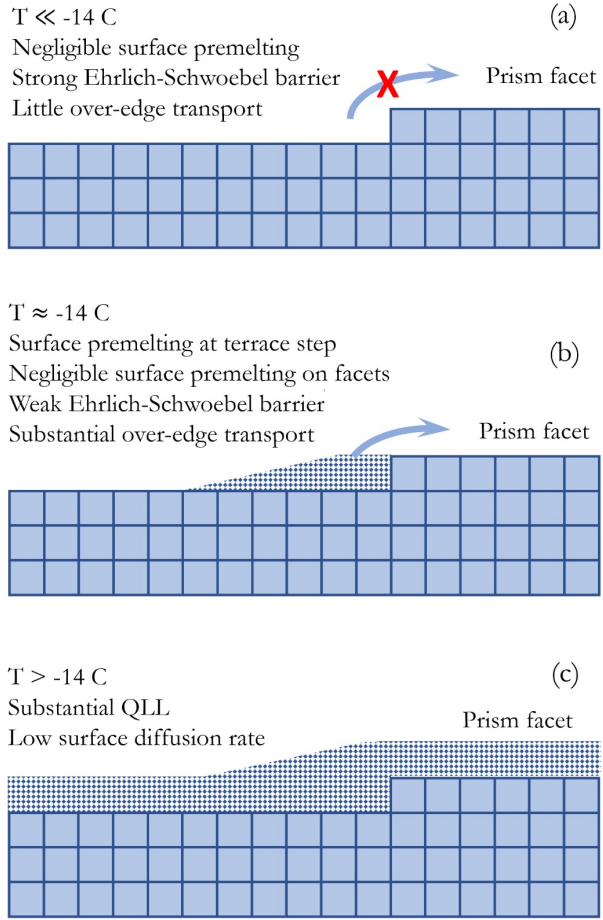

Figure 7: (left) This series of sketches illustrates a possible scenario for the development of surface premelting on a prism facet near -14 C, and how this affects our SDAK model.

(a) When the temperature is far below -14 C, surface premelting is essentially absent and a strong Ehrlich-Schwoebel barrier suppresses admolecule diffusion over terrace steps. In this case, the SDAK effect is absent and $\sigma_{0,prism}$ on a narrow facet is no different than that on a broad facet.

(b) When surface premelting first begins to appear near -14 C, broad prism facets retain their rigid lattice structure, while terrace steps (with weaker molecular binding than broad terraces) exhibit a localized form of premelting. This "softens" the step edge and greatly reduces the Ehrlich-Schwoebel barrier. In this model, surface tension can drive a substantial admolecule flux onto the top prism facet. This yields a large admolecule density perturbation within a distance $x_{SD}$ from the facet edge, facilitating enhanced terrace nucleation.

(c) At higher temperatures, surface premelting covers the entire surface, lowering admolecule diffusion and increasing the admolecule surface density on the top prism facet. As a result, surface tension creates a negligible admolecule perturbation, effectively turning off the SDAK effect.

normal nucleation process and yield a smaller effective $\sigma_{0,prism}$.

Being at the transition between no premelting and substantial premelting, we expect that this phenomenon of localized premelting illustrated in Figure 7b will exist only over a narrow temperature range. Moreover, in this picture, a change in admolecule density occurs at the terrace step, as $\rho_0$ on the upper prism terrace has not yet been substantially affected by premelting. It seems possible, therefore, that this admolecule density jump will result in a higher admolecule transport than that calculated in the simple-surface model, thus substantially enhancing the SDAK effect over a narrow temperature range.

At temperatures far above -14 C, surface premelting occurs over the entire surface, including on broad facets, as illustrated in Figure 7c. In this situation, curvature-driven surface diffusion will occur much as described in the simple-surface model, with the relatively small $x_{SD}$ expected in the QLL, as described above. This results in a negligible SDAK effect compared to that found at the premelting transition temperature.

While it is impossible to calculate such a complex surface behavior as a function of temperature, it seems possible that molecular-dynamics (MD) simulations could address many aspects of this phenomenon. Premelting dynamics have already been studied using these computational tools, even allowing estimates of the terrace step energies on the basal and prism facets as a function of temperature [2020Llo]. As these techniques continue to improve, it may be possible to examine surface diffusion over terrace steps as a function of temperature. Specifically, if MD simulations could examine the Ehrlich-Schwoebel barrier as a function of temperature near the onset of surface premelting, along with surface migration rates under an applied external force, this could potentially provide many insights into the ice attachment kinetics and the SDAK phenomenon.

## ❄ Near-Vacuum Growth

One question that immediately arises in this discussion is why the SDAK effect is seen only in snow crystals grown in air, while experiments performed in a low-pressure environment invariably yield nearly isometric faceted crystals. The primary answer to this question appears to be that the SDAK effect leads to a new kind of diffusion-limited growth phenomenon I have called the edge-sharpening instability (ESI) [2017Lib, 2021Lib]. This mechanism naturally yields thin plates with narrow prism facets near -14 C, but it only operates when the growth is substantially limited by particle diffusion through an inert background gas. The ESI does not occur at low pressures, so thin plates do not appear in near-vacuum conditions. The positive feedback inherent in the ESI drives the formation of thin plates, greatly enhancing the SDAK effect in the process. Because the ESI is not present at low pressure, the SDAK effect cannot becomes as strong as in air.

Another reason is that the SDAK effect happens mainly at high supersaturations that produce corners with small radii of curvature. In these circumstances, the enhanced nucleation can be thought of as an edge effect, as a substantial increase in admolecule surface density will be found only within a distance of roughly $x_{SD}$ from the facet edge. This covers most of the surface on a thin-plate crystal, but only a small fraction of the area on a large faceted surface. Therefore, in near vacuum at fairly high supersaturations, many nucleation events will occur far from the edge region, as the total nucleation rate is proportional to surface area. The interior nucleation will thus tend to mask the SDAK effect.

A final reason is simply that precision ice-growth experiments are quite difficult to perform and are often affected by subtle systematic effects. With improved precision and targeted experiments, it may be possible to see SDAK effects in low-pressure environments. Clearly, performing measurements as a function of background gas pressure will likely yield new insights to the SDAK phenomenon.

## ❄ Conclusions

My goal with this paper is not to provide the final word on snow-crystal attachment kinetics, but rather to continue the discussion and facilitate additional model development, especially regarding the physical origins of the SDAK phenomenon. Hopefully, this will stimulate new investigations that further examine the relevant physical processes, bootstrapping to additional targeted experimental and theoretical studies.

A first conclusion from the investigation described above is that one can quantify surface-tension-driven surface diffusion to some degree, including making "toy" models that identify many of the important physical processes and parameters. Unfortunately, it is a challenge to choose realistic estimates of several quantities, and these often come with significant uncertainties. Nevertheless, our rough estimates clearly suggest that edge-related surface diffusion effects could possibly be substantial in size, perhaps enough to explain the SDAK effect to a large degree.

A second conclusion from this study is that terrace steps play a critical role in the discussion, particularly the final terrace step defining the outline of the top faceted surface. It appears that significant over-step surface transport must occur primarily at this location, resulting in enhanced nucleation only close to this terrace edge. In this sense, the SDAK phenomenon can be thought of as a kind of inhomogeneous nucleation effect, stimulated by surface tension at the edge of a faceted surface.

This result suggests that MD simulations could be applied to better understand the SDAK phenomenon, as the key processes occur mainly at localized terrace steps. Considerable computational and physical accuracy may be needed to reproduce the correct surface-transport effects, but the prospects look good for this general direction of investigation, and many of the computational tools have already been developed.

Clearly, the physics underlying ice growth from water vapor is complex and rather poorly understood at present, but the outlook for making rapid progress looks good. Targeted ice-growth experiments have demonstrated considerable success in facilitating model development, and MD simulations may soon tie the complex molecular processes involved in ice crystal growth to our basic understanding of water molecular interactions.

## ❅ References

[1954Nak] Ukichiro Nakaya, Snow Crystals, Natural and Artificial, Harvard University Press: Cambridge, 1954.

[1957Kob] T. Kobayashi, Experimental researches on the snow crystal habit and growth by means of a diffusion cloud chamber, J. Meteor. Soc. Japan, 35, 71–80, 1957.

[1958Hal] J. Hallett and B. J. Mason, The influence of temperature and supersaturation on the habit of ice crystals grown from the vapour, Proc. Roy. Soc. A247, 440-453, 1958.

[1961Kob] T. Kobayashi, The growth of snow crystals at low supersaturations, Philos. Mag. 6, 1363-1370, 1961.

[1963Mas] B. J. Mason and G. W. Bryant and A. P. Van den Heuvel, The growth habits and surface structure of ice crystals, Phil. Mag. 8, 505-526, 1963.

[1969Bur] J. E. Bertie, H. J. Labbe, and E. Whalley, Absorptivity of Ice I in the Range 4000-30 cm-1, J. Chem. Phys. 50, 4501-4520, 1969.

[1969Fuk] Norihiko Fukuta, Experimental studies on the growth of small ice crystals, J. Atmos. Sci. 26, 522-530, 1969.

[1972Lam] D. Lamb and W. D. Scott, Linear growth rates of ice crystals grown from the vapor phase, J. Cryst. Growth 12, 21-31, 1972.

[1976Rya] B.F. Ryan, E. R. Wishart, and D. E. Shaw, The growth rates and densities of ice crystals between -3 C and -21 C, J. Atmos. Sci. 22, 123-133, 1976.

[1982Bec2] W. Beckmann and R. Lacmann, Interface kinetics of the growth and evaporation of ice single crystal from the vapor phase - Part II: Measurements in a pure water vapour environment, J. Cryst. Growth 58, 433-442, 1982.

[1982Kur] T. Kuroda and R. Lacmann, Growth kinetics of ice from the vapour phase and its growth forms, J. Cryst. Growth 56, 189-205, 1982.

[1983Bec] W. Beckmann and R. Lacmann and A. Blerfreund, Growth rates and habits of ice crystals grown from the vapor phase, J. Phys. Chem. 87, 4142-4146, 1983.

[1984Kur] Toshio Kuroda, Rate determining processes of growth of ice crystals from the vapour phase - Part I: Theoretical considerations, J. Meteor. Soc. Jap. 62, 1-11, 1984.

[1984Kur1] Toshio Kuroda and Takehiko Gonda, Rate determining processes of growth of ice crystals from the vapour phase - Part II: Investigation of surface kinetic processes, J. Meteor. Soc. Jap. 62, 563-572, 1984.

[1989Sei] T. Sei and T. Gonda, The growth mechanism and the habit change of ice crystals growing from the vapor phase, J. Cryst. Growth 94, 697-707, 1989.

[1990Yok] Etsuro Yokoyama and Toshio Kuroda, Pattern formation in growth of snow crystals occurring in the surface kinetic process and the

diffusion process, Phys. Rev. A 41, 2038-2049, 1990.

[1994Ven] J. A. Venables, Atomic processes in crystal growth, Surf. Sci. 299, 798-817, 1994.

[1996Sai] Y. Saito, Statistical Physics of Crystal Growth, World Scientific Books, 1996.

[1998Nel] Jon Nelson and Charles Knight, Snow crystal habit changes explained by layer nucleation, J. Atmos. Sci. 55, 1452-1465, 1998.

[1999Pim] Alberto Pimpinelli and Jacques Villain, Physics of Crystal Growth, Cambridge University Press: Cambridge, 1999.

[1999Pri] W. S. Price, H. Ide and Y. Arata, Self-diffusion of supercooled water to 238 K using PGSE NMR diffusion measurements, J. Phys. Chem. A, 103, 448-450, 1999.

[2001Nel] J. Nelson, Growth mechanisms to explain the primary and secondary habits of snow crystals, Phil. Mag. 81, 2337-2373, 2001.

[2001Tso] Tien T. Tsong, Mechanisms of surface diffusion, Prog. Surf. Sci. 67, 235-248, 2001.

[2002Mut] Boyan Mutaftschiev, The Atomistic Nature of Crystal Growth, Springer-Verlag: Berlin, 2002.

[2003Lib] K. G. Libbrecht, Growth rates of the principal facets of ice between -10C and -40C, J. Cryst. Growth 247, 530-535, 2003.

[2003Lib1] K. G. Libbrecht, Explaining the formation of thin ice-crystal plates with structure-dependent attachment kinetics, J. Cryst. Growth 258, 168-175, 2003.

[2004Bai] M. Bailey and J. Hallett, Growth rates and habits of ice crystals between -20C and -70C, J. Atmos. Sci. 61, 514-544, 2004.

[2009Bai] Matthew Bailey and John Hallett, A comprehensive habit diagram for atmospheric ice crystals: confirmation from the laboratory, AIRS II, and other field studies, J. Atmos. Sci. 66, 2888-2899, 2009.

[2011Gla] Ivan Gladich et al, Arrhenius analysis of anisotropic surface self-diffusion on the prismatic facet of ice, Phys. Chem. Chem. Phys. 13, 19960-19969, 2011.

[2012Kni] Charles A. Knight, Ice growth from the vapor at -5 C, J. Atmos. Sci. 69, 2031-2040, 2012.

[2012Lib2] K. G. Libbrecht, On the equilibrium shape of an ice crystal, arXiv:1205.1452, 2012.

[2013Lib] Kenneth G. Libbrecht and Mark E. Rickerby, Measurements of surface attachment kinetics for faceted ice crystal growth, J. Crystal Growth 377, 1-8, 2013. Preprint at arXiv:1208.5982.

[2015Lib] Kenneth G. Libbrecht, The surface diffusion length of water molecules on faceted ice: A reanalysis of 'Roles of surface/volume diffusion in the growth kinetics of elementary spiral steps on ice basal faces grown from water vapor' by Asakawa et al., arXiv:1509.06609, 2015.

[2017Lib] Kenneth G. Libbrecht, Physical dynamics of ice crystal growth, Ann. Rev. Mater. Res. 47, 271-95, 2017.

[2019Lib1] Kenneth G. Libbrecht, A quantitative physical model of the snow crystal morphology diagram, arXiv:1910.09067, 2019.

[2019Lib2] Kenneth G. Libbrecht, Toward a comprehensive model of snow crystal growth: 6. Ice attachment kinetics near -5 C, arXiv:1912.03230, 2019.

[2020Lib1] Kenneth G. Libbrecht, Toward a Comprehensive Model of Snow Crystal Growth: 8. Characterizing Structure-Dependent Attachment Kinetics near -14 C, arXiv:2009.08404, 2020.

[2020Lib2] Kenneth G. Libbrecht, Toward a Comprehensive Model of Snow Crystal Growth: 9. Characterizing Structure-Dependent Attachment Kinetics near -4 C, arXiv:2011.02353, 2020.

[2020Llo] Pablo Llombart, Eva G. Noya, and Luis G. MacDowell, Surface phase transitions and crystal growth rates of ice in the atmosphere, Science Advances 6, no. 21, eaay9322, DOI:

10.1126/sciadv.aay9322, 2020. Preprint available at arXiv:2004.10465, 2020.

[2021Lib] Kenneth G. Libbrecht, Snow Crystals, Princeton University Press, in production, to appear in 2021.